\documentclass[12pt]{iopart}
\usepackage{graphicx}

\begin{document}

\title[Statistics of the excited spectra of the Jahn-Teller
system]{Marginal intermediate statistics in the excited spectra of
E$\otimes (b_1+b_2)$ Jahn-Teller system}

\author{E Majern\'{\i}kov\'a$^{1,2}$ and S Shpyrko$^2$
\footnote{On leave from the Institute for Nuclear Research,
Ukrainian Academy of Sciences, pr. Nauki 47, Kiev, Ukraine; email
serge\_shp(at)yahoo.com}}

\address{$^{1}$Institute of Physics, Slovak Academy of Sciences,
D\'ubravsk\'a cesta 9, SK-84 511 Bratislava, Slovak Republic }
\address{$^2$Department of Optics, Palack\'y University, T\v r. 17.
listopadu 50, CZ-77207 Olomouc, Czech Republic}

\ead{Eva.Majernikova@savba.sk}

\begin{abstract}
 The long-range spectral density correlations (spectral
rigidities $\bar{\Delta}_3(\bar n)$ and related spectral
compressibilities) of the $E\otimes (b_1+b_2)$ Jahn-Teller model are
found strongly nonuniversal with respect to the Hamiltonian
parameters and inhomogeneous with respect to the choice of a partial
energy segment.
 However, the partial spectral rigidities exhibit common features:
 an anomalous linear part for small  $\bar n$ and a saturation for large $\bar n$.
The spectral compressibilities  found for the partial spectral
segments and averaged over a whole relevant part of the spectrum
cumulate close to a well-defined limit pertaining to the
semi-Poisson statistics. This is in accordance with similar tendencies
revealed in the short-range averaged statistical characteristics of
this model investigated in our previous paper (Majern\'{\i}kov\'a and  Shpyrko 
Phys. Rev. E{\bf 73} 057202).
These features document an inhomogeneous and nonuniversal weakening of
level repulsions  and nonuniversality of level fluctuations on both
long and short energy scales. The nonuniversality and inhomogeneity  of the
statistical characteristics correspond to a similar behaviour of the
chaoticity parameter (a fraction of the chaotic phase space of the
trajectories) found for the corresponding semiclassical Hamiltonian.
We ascribe the nonuniversal and inhomogeneous nonintegrability
behaviour to the changing degree of the brokening of the  rotation
symmetry when changing parameters of our effectively two-dimensional
model. It results in a random distribution of the respective
localized wave functions at all scales up to the size of an
available state space. The multifractal behaviour of the wave
functions is implied from the analysis of their (averaged) fractal
dimensions which range up to $1.5\pm 0.1$ (for $\bar{D}_2$). This
might imply the concept of the chaos-assisted tunnelling between the
regions of reduced degree of stochasticity through regions of high
degree of stochasticity.  It supports the analogy with the
two-dimensional Anderson model with marginal-asymptotically far
metal-insulator transition. The features found  allow us to classify
the present model as a member of the class with a multifractal
eigenfunction statistics characteristic for the spectra with
weakened level repulsion similar to the Anderson model close to
 the metal-insulator transition.

\end{abstract}


\maketitle

\section{Introduction}
Investigations of a wide class of complex Hamiltonians  show a
coexistence of ordered and more disconnected chaotic dynamic regions
in their phase space in a semiclassical limit \cite{Bohigas:1993}.
Respective parameters, fractions of the regular and chaotic regions
of the phase space were used for the evaluation of statistical
characteristics of the related quantum spectra. Namely, explicit
formulae for the nearest-neighbour level spacing distributions
\cite{Berry:1984} and the averaged level number variances or the
$\Delta_3$ statistics \cite{Seligman:1985a,Seligman:1985b} were
illustrated by a model of two coupled quartic oscillators. The level
repulsion at small level spacings was shown to be reduced by the
fractionating of the chaotic region which substantially complicated
related numerical calculations (level spacing distributions).
Discrepancies between the numerical data and the analytical formulae
for the above mentioned short-range and also the middle and 
long-range fluctuation statistics were ascribed especially to quantum
effects like the tunnelling between regular regions through chaotic
barriers and the quantum interference effects.
 Finally, the chaos-assisted tunneling approach
to the calculation of the level repulsion implied the correct
behaviour of the level spacing distributions in the limit of small
level separations \cite{Podolskiy:2003}. This kind of tunnelling was
shown to be connected with the fractal dimension of a
two-dimensional phase space due to a hierarchical phase space
structure of a chaotic system \cite{Ketzmerick:1996}.

In this paper, we try to show that one of the systems which exhibit
fluctuation properties typical for a transition region between order
and chaos with varying (nonuniversal) degree of nonintegrability is
the $E\otimes (b_1+b_2)$ Jahn-Teller (JT) model. The model is
represented by  two degenerate electron levels coupled to two phonon
(vibron) modes via two different interaction strengths. Unitary
diagonalization of the Hamiltonian in the electron space transforms
the system onto two highly nonlinearly coupled quantum oscillators
in two dimensions.

 In a recent paper \cite{Majernikova:2006b}, we have initiated the study
 of the statistical evidence of quantum chaotic patterns emerging in this
model. We have shown that the (short-range) statistics of the
nearest-neighbour level spacings (NNS) in the range of  interaction
parameters apart from the particular symmetry cases (E$\otimes$e JT
and exciton model) is nonuniversal and tends to a well-defined limit
 close to the semi-Poisson law $P(S)=4S \exp
(-2S)$.  This intermediate statistics between the Poisson (uncorrelated
levels) and the Wigner-Dyson distribution of a correlated fully chaotic
level system was found within the frame of random matrix theory as a
critical distribution in the metal-insulator transition region in
the Anderson model of disorder \cite{Shklovskii:93,Evangelou:00} as
well as in several models with not fully developed chaotic dynamics
by Bogomolny \cite{Bogomolny:1999}. Namely, the semi-Poisson
distribution of  the level spacings
 in the plasma model was ascribed to the screening (restriction to a
finite number of nearest neighbours) of the logarithmic pair
interaction potential. Statistical methods for energy levels and
eigenfunctions of disordered systems were reviewed by several
authors \cite{Mirlin:2000}.

In section 2, we shall concentrate on the complementary item of
the long-range statistics for the excited spectra of the $E\otimes
(b_1+b_2)$ JT model with broken rotation symmetry. It comprises the
spectral rigidity measure of correlations of the level density on
scales large when compared to the mean level spacings. Among
possible variants of this measure one can cite the $\bar{\Delta}_3 $
statistics of Dyson and Mehta
\cite{Shklovskii:93,Evangelou:00,Dyson:4,Mehta:1960}. Namely, the 
$\Delta_3$ value
is defined as a (random) quantity which for a given energy interval
$[-L,L]$ around the value $E_0$ gives the deviation of the
least-square linear fit line from the staircase function $N(E)$ of
the number of levels with energy below $E$:

\begin{equation}
\Delta_3(L)= {\textrm {min}}_{A,B} \left\{\frac{1}{2L}
\int\limits_{-L}^{L} [N(E)-AE-B]^2\mathrm{d}E \right\}
\end{equation}
(here the energy is shifted so that the interval is centered around
$0$; the minimalization is performed with respect to parameters
$A,B$). The average of this measure over ensemble $\bar{\Delta_3}$
is the quantity of the present interest. The $\bar{\Delta_3}$
measure and the correlation function of the level number
$\Sigma_2(\bar{n})\equiv \langle \delta^2 N\rangle= \langle
N^2\rangle - \langle N \rangle^2$  (fluctuations of the level number
in an energy band of a width $E$, $\bar n\equiv \langle N \rangle$
is the length of the energy interval measured by the mean number of
the levels inside) are  related  straightforward as
$\bar{\Delta_3}(\bar{n})=\frac{2}{\bar{n}^4}\int_0^{\bar{n}} \left(
\bar{n}^3 - 2 \bar{n}^2 r + r^3 \right) \Sigma_2(r) \mathrm{d}r $
\cite{Pandey:3}. The use of the $\bar{\Delta_3}$-statistics instead
of the level number fluctuations $\Sigma_2$ is justified by the fact
that its variance is suppressed when compared to that of $\langle
\delta^2 N\rangle$ \cite{Dyson:4}.

The random matrix theory (RMT) predicts the scaling of this measure
as $\log \bar n$ (or $\log E $) for the domain of
 fully developed chaos, meanwhile for the completely uncorrelated
sequences of levels a linear scaling $\sim \bar n$ is expected
\cite{Bogomolny:1999,Kravtsov:1995}.

For the Poisson ensemble of an uncorrelated
sequence of levels one has  $\Sigma_2(r)=r$ and $\bar{\Delta}_3(\bar
n)=\bar{n}/15$. The number variance $\Sigma_2(\bar n)$ was
calculated for a set of ensembles within RMT supposed to model the
behaviour of Anderson type systems
\cite{Kravtsov:1995,Chalker:1996}. Its asymptotic form was shown to
be similar to the uncorrelated case but with the coefficient (level
compressibility) $\chi<1$, $\Sigma_2(\bar n)\sim \chi \bar n $ for
large $\bar n$. The value $\chi >0$ at weakened level repulsion
(compared to the metallic limit) refers to the fractal nature of
wave functions.
 Restricting to the nearest neighbour interactions of the logarithmic pair
potential in the Coulomb plasma model Bogomolny {\it et al} \cite{Bogomolny:1999}
found the value of the compressibility $\chi=1/2$ which should hold for the
models with the semi-Poisson statistics.

In section 3, we discuss a completing semiclassical concept of
the degree of chaoticity, i.e., of  the  phase space fraction
occupied by the chaotic trajectories $\mu$ for different segments of
the spectra and interaction parameters. A broad region of $\mu$ from
increasing to maximum values in the low and middle parts of the
spectra to the  decreasing again values in the upper parts of the
spectra occurs, strongly depending on the parameters and location of
the spectral segment. These results confirm the nonuniversality and
varying nonintegrability properties as obtained in section 2.

In section 4, an independent quantitative statistical analysis
of quantum chaotic patterns of long-range type is provided by the
fractal properties of the eigenfunctions: the set of fractal
dimensions $D_q$ is defined by the scaling properties of the inverse
participation ratios $I_q(n)$ related to the eigensolution of the
state $n$ of a Hamiltonian as $I_q(n)= \sum\limits_r |\Psi_n
(r)|^{2q}$. For the sake of application to the excited states of our
electron-vibron model (defined in section 2) it is suitable to
use the spectral representation where the relevant substrate is the
space spanned by the vibron (phonon) Fock states $n$
\cite{Majernikova:2006a}
\begin{equation}
I_q (n) \equiv \sum_i |C_{in}|^{2q}. \label{D}
\end{equation}
 Here, $C_{in}\equiv \langle \Phi_i | \chi_n \rangle$; $\chi_n(Q_1, Q_2)$ are the exact
 (numerically calculated for the present model)
vibron  wavefunctions in the space of coordinates $Q_1$ and $Q_2$; the base set $\Phi_i$
is chosen as the set of vibron Fock states - excited  states $i$ of the unperturbed
two-dimensional harmonic oscillator, namely, $i$ is a compound index denoting the
direct product of the excited states $s$ and $t$ of the vibrons $1$ and $2$:
$|i\rangle \equiv |s\rangle_1 \otimes |t\rangle_2
\sim b_1^{\dag s} b_2^{\dag t} |0\rangle_1 \otimes |0\rangle_2$.

  The scaling with the
fractal dimension $D_q$ in the spectral representation assumes that
one explores the probabilities $P_{i,L}(n) = \sum_{s\in i}
P_{s}(n)=\sum_s C_{s\in i, n}^2$, where the sum is taken over the
states inside the cube $i$ of dimension $2$ comprising $l^2$ base
Fock states and $L$ denotes the number of these cubes ($L\sim
1/l^2$). Then,
\begin{equation}
I_{q, L} (n) \propto
\sum_{i}^{L}P_{i, L}^{q}(n)  \propto
L^{-D_q(q-1)} \,,\label{IPR}
\end{equation}
where the information dimension $D_1$ is understood as usual in the
limit $q \to 1$ as the scaling factor of $\exp(-\sum P_i \log P_i)
\propto L^{D_1}$. A common assertion  found by Kravtsov and Muttalib
\cite{Kravtsov:1997} states that the critical statistics in certain
class of systems including Anderson model at M-I transition is in an
intimate relation with the weakly overlapping (implying
multifractality) wavefunctions.

\section{ $\bar{\Delta}_3 $ statistics as a measure of spectral
rigidity of the E$\otimes(b_1+b_2)$ Jahn-Teller model}

The  E$\otimes(b_1+b_2)$ JT model is defined by the Hamiltonian

\begin{equation}
\hat{H}= \Omega (b_{1}^{\dag}b_{1} +b_{2}^{\dag}b_{2}+1 )I + \alpha
(b_{1}^{\dag}+b_{1})\sigma_{z}
 -\beta (b_{2}^{\dag}+b_{2})\sigma_{x}
\label{H:init}
\end{equation}
described, e.g., in our previous papers
\cite{Majernikova:2006b,Majernikova:2006a,Majernikova:2003}.
 In short, the local spinless double
degenerate electron level is linearly coupled to two intramolecular
vibron (phonon) modes of the frequency $\Omega$ by different
coupling constants $\alpha\neq \beta$.  The pseudospin notation with
$2\times 2$  Pauli matrices $\sigma_x$, $\sigma_z$ and unit matrix
$I$ refers to the two-level electron system. The operators $b_i$,
$b_i^{\dag}$ satisfy boson commutation rules
$[b_i,b_j^{\dag}]=\delta_{ij}$ and define the vibron coordinates
$Q_i= \langle b_i^{\dag}+b_i\rangle $, $i=1,2$. The interaction term
$\propto\alpha$ removes the degeneracy of the electron levels and the
term $\propto\beta$ mediates the vibron-assisted electron tunnelling
between the levels (the class of exciton models differs from the
present one by the absence of the vibron-2 assistance in the
tunnelling term  $\sim \beta$). Two cases of special symmetry
comprise the rotation symmetric E$\otimes$e JT model
($\alpha=\beta$) and the polaron model with $\beta=0$ or $\alpha=0$.
Introducing nonequal coupling constants the model above
generalizes the common E$\otimes$e Jahn-Teller model. The interest for such
a generalization has various sources, in particular: (a) different
coupling constants are likely to be caused by a spatial anisotropy
in a crystal plane; (b) two vibron modes need not have necessary the
same frequencies. In this case, the Hamiltonian can be appropriately
rescaled to the present model with equal frequencies but different
coupling strengths (in what follows we set $\Omega=1$); and, last
but not least, (c) the generalized JT system with broken rotational
symmetry presents  more rich variety of generic chaotic properties
than its simpler prototype. As it was shown
\cite{Majernikova:2006b,Majernikova:2006a}  it stands closer to the
generic models of quantum chaotic behaviour.

In the following we use the vibron eigenfunctions of the transformed
Hamiltonian $\tilde{H}\equiv U \hat H U^{-1}$
\cite{Majernikova:2006a,Majernikova:2003}
\begin{equation}
\tilde{H}=
 \Omega (\sum_{i=1,2}b_{i}^{\dag}b_{i}+1 ) +\alpha
(b_{1}^{\dag}+b_{1}) -p\beta (b_{2}^{\dag}+b_{2}) R_{ph}\, \label{H}
\end{equation}
exactly diagonalized in the electron subspace by the
Fulton-Gouterman (FG) unitary operator $ U= \frac{1}{\sqrt 2} \left (
\matrix{1\ , \ R_{ph}\cr  1\ , \ -R_{ph}}\right ), $ where
$R_{ph}=\exp(i\pi b_{1}^{\dag}b_{1})$ is the vibron reflection
operator imposing high nonlinearity in the system and the parity
$p=\pm 1$. It is easy to see
that in the FG representation the parities $p=\pm 1$ are exactly
mirror images of each other and can be mapped one onto
another by a mere change of the sign in the definition of the phonon-2
displacements ($\hat{Q_2} \to -\hat{Q_2}$), thus the system remains
doubly degenerated and the parity does not have any impact on the
properties of the spectrum; in  what follows we choose $p=+1$.

\begin{figure}[htb]
\includegraphics[width=0.9\hsize]
{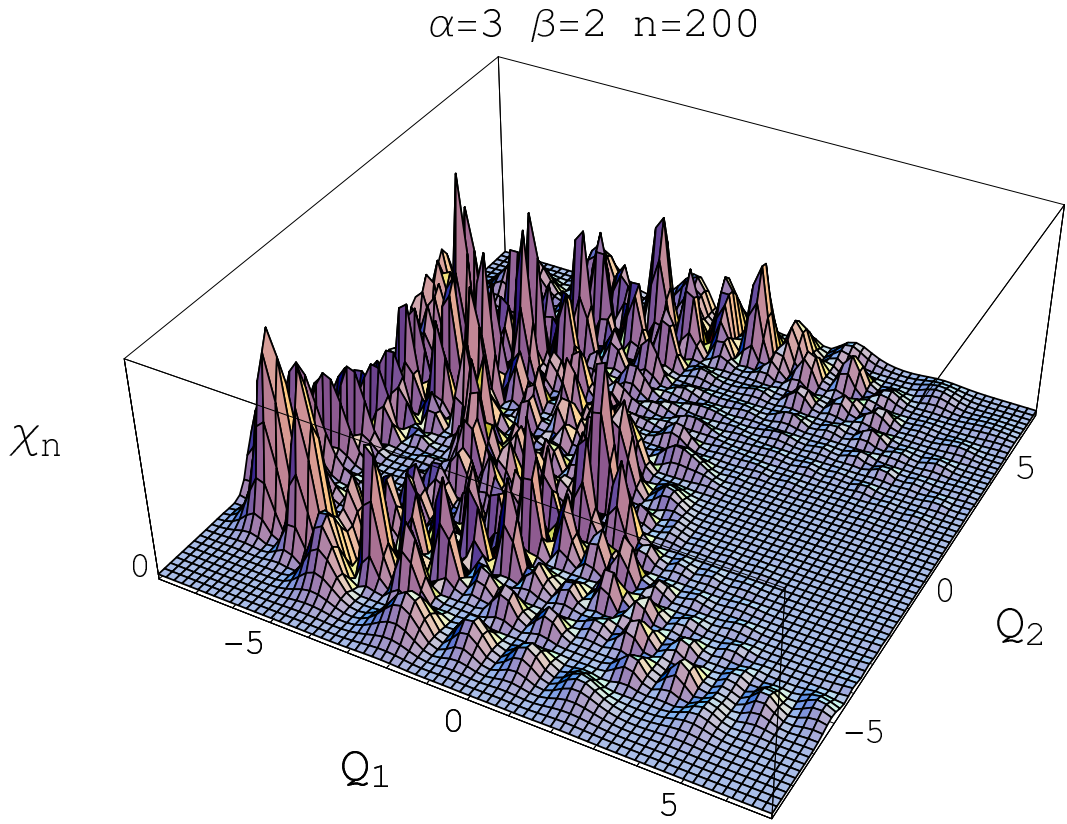} \caption{Example of the wavefunction
$\chi_{n}(Q_1, Q_2)$ for $\alpha=3$, $\beta=2$ and $n=200$.}
\label{fig:1}
\end{figure}

 In the representation
of radial coordinates in the plane $Q_1 \times Q_2$,
$Q_1=r\cos\phi,\ Q_2=r\sin \phi $ the transformed Hamiltonian
(\ref{H}) yields \cite{Majernikova:2006a}
\begin{eqnarray}
\tilde{H}= -\frac{1}{2r}\frac{\partial}{\partial r} \left(
r\frac{\partial}{\partial r}\right) +\frac{1}{2}r^2-\frac{1}{2
r^2}\cdot \frac{\partial^2}{\partial \phi^2}\nonumber\\
+\sqrt{2}\alpha r (\cos\phi -\sin\phi R_{ph} )I +p
\sqrt{2}(\alpha-\beta)r \sin\phi R_{ph}  \,. \label{Ham}
\end{eqnarray}
The reflection operator $R_{ph}$ in radial coordinates
acts as $R_{ph}(r,\phi)f(r,\phi)=f(r,\pi-\phi)$ on some $f(r,\phi)$.
 An example of the
numerical eigenfunction to the transformed Hamiltonian (\ref{H}) in
the space $Q_1\times Q_2$ is shown in figure \ref{fig:1}. Let us note the
apparent fractal nature of the space distribution of the state
amplitude\footnote{It is to be noted that the energy spectrum of the system is
invariant with respect to the interchange $\alpha \leftrightarrow
\beta$. But the system itself is not invariant with respect to this
interchange. For example, the ground state is essentially different in
the domains of heavy ($\alpha>\beta$) and light ($\alpha < \beta$)
polarons \cite{Majernikova:2003}. However, in the Fock state representation the
components of the corresponding wave vectors differ up to the sign
change which does not affect the fractal properties of wavefunctions
in the spectral
representation and, hence, the symmetry of exposed results with
respect to the said transformation.}. The Hamiltonian (\ref{Ham}) commutes with the
operator of the angular momentum $\hat{J}=i (b_1 b_2^{\dag}-
b_1^{\dag}b_2) -\sigma_y/2$ if $\alpha=\beta$. Thus, the
eigenfunctions of the symmetric model can be chosen each to pertain
to a state with a good quantum number $|j|=1/2, 3/2, \dots$
\cite{Majernikova:2006a}.

\begin{figure}[htb]
\includegraphics[width=0.9\hsize]
{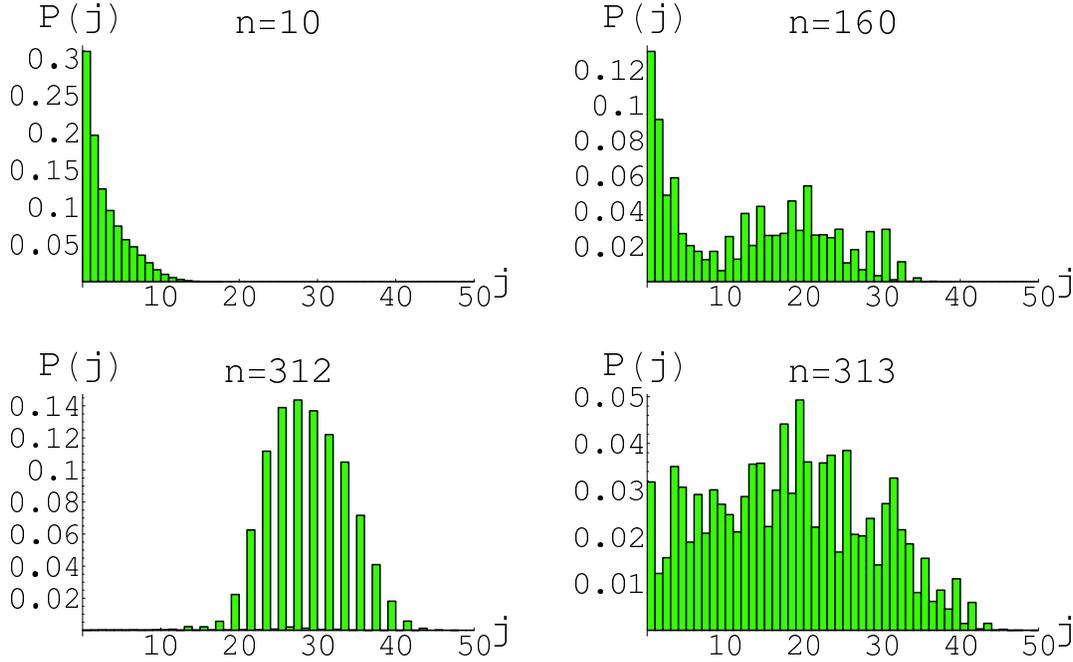} \caption{Examples of distributions of wavefunctions over $j$-space for $\alpha=2, \beta=3$. Evidently, the
states exhibit different extent of localization.}
\label{fig:2}
\end{figure}

 The last term in (\ref{Ham})
 breaks the rotational symmetry and involves
interactions of the basis states with different $j$. Hence, a
wavefunction of the full Hamiltonian is now distributed over a range
of values of the angular momentum (figure \ref{fig:2}). An analogy to the
two-dimensional Anderson  model of disorder can be traced if the
base functions with definite $j$ are considered as 'pseudosites'
over which the wavefunction is spread. Similarly as in the Anderson
model, the properties of the model are determined by the relative
strengths of the intersite (different $j$) and the onsite
interactions. The microscopic  reason for this analogy is the
assistance of phonons-2 in transitions between the levels
(\ref{Ham}) which allows for the emerging of the pseudolattice. To
our opinion, it is just these transitions with changing symmetry
($R_{ph}$) which ensured the similarity of the results for NNS
distribution \cite{Majernikova:2006b} to the Anderson model close to
the M-I transition (marginal in two-dimensions). Traces of quantum
chaos cause the sequence of these coefficients to be distributed in
a random fashion with varying $n$ thus supporting the analogy to the
Anderson model with {\it a priori} given random coefficients.

Calculating the angular part of the matrix elements of the symmetry
breaking term in (\ref{Ham}) between the states of harmonic
oscillator the selection rules can be obtained which assert that the
element with definite $j$ is directly connected to the elements with
$j'=j+2,j-2$ (cf.especially figure 2 with $n=312$ where only every
odd `pseudosite' $j$ is populated). This reminds us of an analogy with the `daisy
models' of random matrix ensembles \cite{Hernandez:1999} with
dropping every second term which were shown to lead directly to the
semi-Poisson distribution of NNS. The natural 'length' of our
pseudolattice (for a given interval of energy) is thus the maximal
number of allowed values of angular momentum for a given n (level
number), that is $L \sim \sqrt{n}$ (a state with main quantum number
$n_r=0,1,\dots$ of a two-dimensional harmonic oscillator is $n_r+1$
times degenerated with auxiliary quantum number ranging between $0$
and $n_r$, hence $n\sim n_r^2$). Thus, the analogy to the Anderson
type models with M-I transition whose intrinsic characteristics is
the length  $L$ of a system appears now more pronounced.

In order to calculate the long-range averaged characteristics of the
energy spectra of (\ref{H}) we numerically diagonalized the
Hamiltonian matrix in the representation of the base Fock states of
the vibrons $1$ and $2$. Taking $N_1$ and $N_2$ base Fock states for
each vibron we introduced an appropriate ordering of sites
\cite{Zyczkowski:1994} so that a state vector turned to be a vector
with $N_1\cdot N_2$ elements. Numerical determination of high energy
states produces an inevitable error because of truncating the base
space. Varying the numbers of the base states we investigated the
convergence of the results. For practical purposes we limited
ourselves by the base size $N_1\times N_2=75\times 75$ from whence
about 1100-1200 states appeared to be trusty (giving the convergence
up to 0.1$\times$(characteristic level spacing)). This number of
energy levels was used in the following calculations. Thus, the obtained
spectrum needs to be unfolded in order to ensure its homogeneity
(that is normalized so that the local averaged level spacing is
unity). To do so it is necessary to fit the 'staircase' $N(E)$ (or
the corresponding level density $\rho(E)=\sum_i\delta(E-E_i)$) by a
reasonably chosen smooth function. Different choices of the latter
are possible. The calculations below used the unfolding by fitting
the level density by a third-order spline polynomial (after
smoothing the data via Gaussian smoothing with the smoothing
constant $0.3$) in the domain of interest, that is for level
numbers $100-1200$. In this interval the level density of our
two-vibron system grows almost linearly with the energy (we
note by passing that this behaviour is in a qualitative accordance
with the semiclassical Weyl formula giving generically a linear
growth of the level density for a two-vibron
system), so that such fitting turns to be more than satisfactory. We
performed sample calculations with an increasing order of polynomials,
and using other algorithms of producing the smoothed level density
$\rho_{sm}$ as well (in particular, varying the parameters of the
Gaussian smoothing and using the smoothing via moving average/median);
the results appeared indistinguishable within statistical errors.

In order to compile a statistical ensemble out of the
(deterministic) quantum problem one has to perform averaging of the
numerical quantities of interest over different parts of the
spectrum taken as members of this ensemble.  We included into the
statistical analysis the energy intervals centered at different
starting values ranging between 100-1100 levels which corresponded
to the energy values between $(-10)$ and $(-5)$ and between $25$ and $30$
depending on the values of ($\alpha,\beta$). However, as it follows
from the results below, the ensemble averaged properties can show a
bias (sensitivity to the starting point in the energy range),
therefore this procedure must be followed with maximal care. To
improve the statistics we also followed the standard procedure of
collecting statistical data from small intervals in the space of
parameters $(\alpha, \beta)$. To be exact, for a given point of
interest $(\alpha_0, \beta_0)$ we collected the energy values from
the square $(\alpha_0,\beta_0) \pm 0.25$ in the parameter space with
the step $0.1$. (The exception of this rule was collecting the data
for the points where $\alpha_0=\beta_0$. There, this restriction was
followed for choosing the neighboring parameters for the sampling).
From figure 4 it follows that indeed the averaged statistical data
present a slight bias when taking different energy intervals.
Therefore we presented the results in figure \ref{fig:4} separately in two
energy bins for lower (100-600) and higher (600-1100) levels. It is
worth noting that the nonsymmetric parameter values ($\alpha\neq \beta$)
appear to present more robust statistics with respect to the energy
interval than the symmetric ones ($\alpha=\beta$). The same
conclusion was already drawn from the investigation of the
short-range statistics \cite{Majernikova:2006b,Majernikova:2006a}
where we showed that the spectral properties of the nonsymmetric JT
models were more homogeneous with respect to the choice of the
energy interval. Taking smaller energy bins than those of figure 4 however
did not change much the appearance of the corresponding figures.

\begin{figure}[h]
\includegraphics[width=0.9\hsize]
{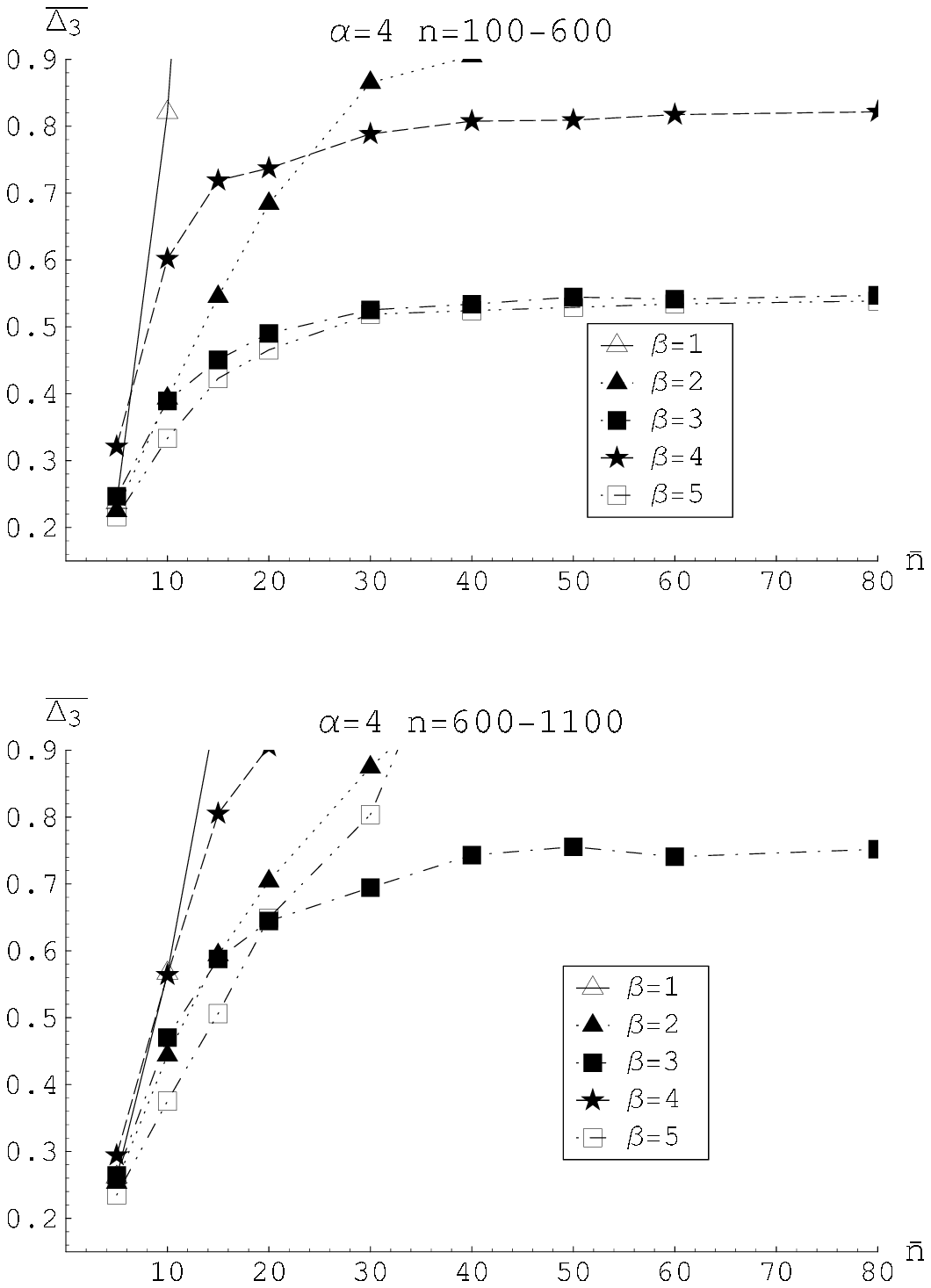} \caption{Spectral rigidities ($
\bar{\Delta}_3$-statistics) in different segments of the spectra for
the E$\otimes(b_1+b_2)$ JT model. The slopes are linear  for $\bar
n$ at least  up to 15 and tend to nonuniversal saturation values
$\bar \Delta_{3 max}(\beta)$ for large $ \bar{n}$.}
\label{fig:3}
\end{figure}

In figure \ref{fig:3}, the samples of the long-range spectral correlations
 $\bar{\Delta_3}$  of the $U$-transformed
 Hamiltonian (\ref{H}) (unfolded in a fashion described above)
 are presented for several sets of interaction
parameters. It is seen that the spectral rigidity shows up serious
deviations from the Poisson behaviour as well as from the
logarithmic dependence expected in the frame of RMT. For large $\bar
n$ it supposedly tends to a saturation, and a characteristic linear
domain is evident for small $\bar n$ (up to $\bar{n}\sim 15$).

\begin{figure}[h]
\includegraphics[width=0.6\hsize]
{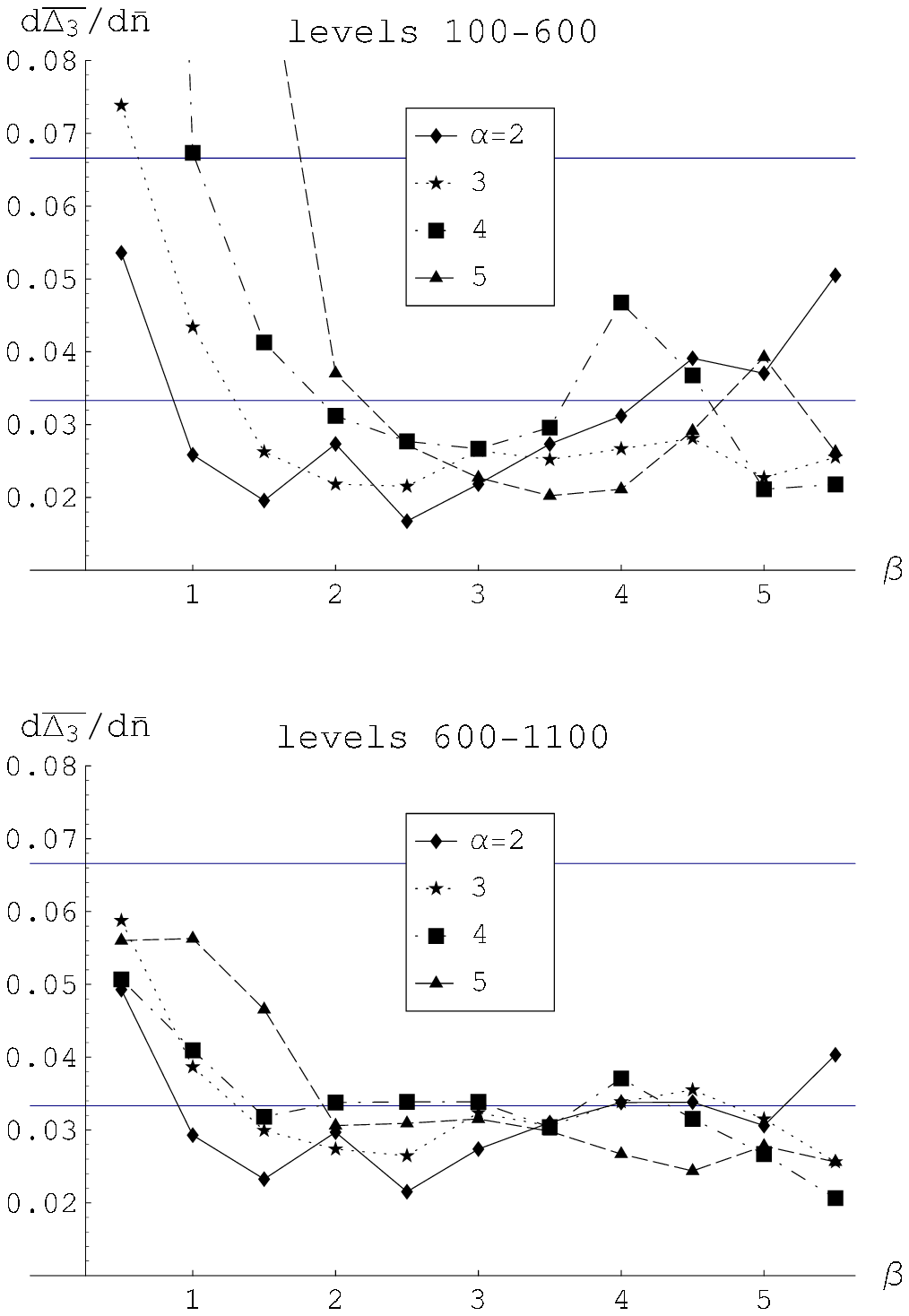} \caption{Slopes
$\mathrm{d}\bar{\Delta}_3(\bar{n})/\mathrm{d}\bar{n}_{|\bar{n}=0}$
('level compressibilities') of the long-range statistical measure
for different model parameters $(\alpha,\beta)$ for different parts
of the spectra. The grid lines indicate the Poisson line (1/15) and
the marginal semi-Poisson limit ($\chi= 0.5/15\simeq 0.033$). The
dispersion of the probability distributions pertaining to the
displayed mean values is $\sim 0.01$. The pictures show a strong
dependence on the position of the fragment within the spectra.}
\label{fig:4}
\end{figure}

\begin{figure}[h]
\includegraphics[width=0.8\hsize]
{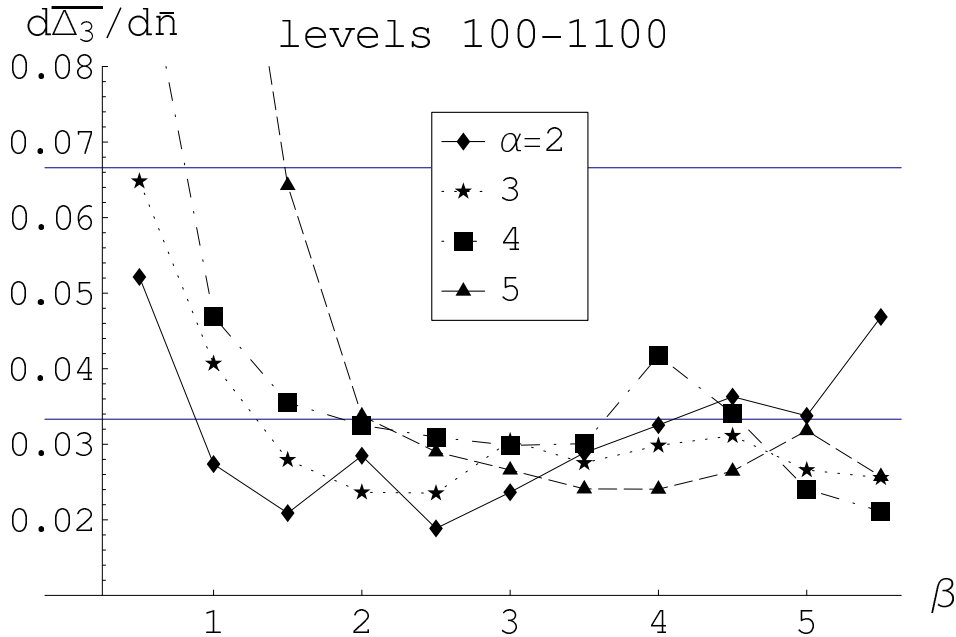} \caption{The cumulative slopes
$\mathrm{d}\bar{\Delta}_3(\bar{n})/\mathrm{d}\bar{n}_{|\bar{n}=0}$
 averaged over the relevant extent of the spectra.  Large deviation of the
points  $\alpha=4$, $\beta>4$ from the suggested marginal line is
explained by their relative proximity to the symmetric point
$\alpha=\beta=4$. }
\label{fig:5}
\end{figure}

In figures \ref{fig:4} and \ref{fig:5}, we plot the level compressibilities (slopes $\mathrm{d}\bar{\Delta}_3(\bar n)/\mathrm{d}{\bar n}_{|\bar n=0}$ of
the $\bar{\Delta}_3$ curves) for a range of parameters $\alpha,
\beta$. The slopes in figures \ref{fig:4}a, \ref{fig:4}b were calculated from averaging over 500 levels in the ranges 100-600 and 600-1100, respectively.
Taking lower levels would bring us below the 'diabatic line' where
the system feels only one potential well of the 'effective
potential' (see \cite{Majernikova:2003,Majernikova:2006a} for a
detailed discussion). It is seen from figure 4 that results for
different intervals of energy show differences, but, nevertheless,
similar tendencies. Apart from the cases close to the special
symmetries ($\alpha=\beta$ and $\alpha \ll \beta$, $\alpha \gg
\beta$) the slopes show a markable accumulation close to the value
$\chi=0.5/15$ (the horizontal grid line at 0.033) indicating the
semi-Poisson limit. This limit is more pronounced at higher levels
(interval 600-1100), where even the points corresponding to the
symmetric cases $\alpha=\beta$ are close to this line. For lower
energy interval (levels 100-600) this universality is seen less,
although all $\bar{\Delta_3}$ curves points are far from both
Poisson and RMT limits. For lower energies there is also a more
pronounced difference between symmetric and nonsymmetric models:
meanwhile the nonsymmetric values of parameters still tend to the
semi-Poisson limit, the symmetric points however show essential
deviation from this limit towards the Poisson case.

In figure \ref{fig:3}, the $\bar{\Delta}_3$ curves for $\beta=2,4$  for
$\alpha=3$ (scaled by $\Omega=1$) averaged over the whole relevant
extent of the spectra (up to the level $n=1100$) are situated very
close one to another, i.e. they are getting close to a limit with
the minimal slope.

 The said marginality also emerged in the statistics of NNS
\cite{Majernikova:2006b}. We have shown that the dispersions
$\sigma^2$ of NNS distributions in the range of parameters far
from the mentioned special symmetry cases tend to a
 limit $\sigma^2\simeq 0.5$ characteristic for the M-I transition of
the Anderson model.

\section{Quasiclassical description in the phase space}

It is interesting to discuss the mapping between the exposed results
and the semiclassical description of the same system. The search for
the quantum chaotic patterns historically meant looking for quantum
correspondence to the chaotic behaviour of the classical
trajectories. First of all it is to be noted that the passage to the
(semi)classical description in a two- or many-level electron-phonon
system can be performed in several fashions. The most
straightforward "semiclassical" description of the quantum JT
problem is obtained if we pass to the real Bloch variables made of
pseudospin Pauli matrices:

\begin{equation}
x(t):=\langle\sigma_x \rangle_t\label{29a} \quad
y(t):=\langle\sigma_y \rangle_t \label{30a} \quad
z(t):=\langle\sigma_z \rangle_t\label{31a}
\end{equation}
and perform the appropriate decoupling of the oscillatory (phonon)
and electron variables in the initial Hamiltonian (\ref{H:init})
within the framework of the Born-Oppenheimer approximation: $\langle
\hat{Q_1} \hat{\sigma_x} \rangle \to \langle\hat{Q_1}\rangle(t)
x(t)$ etc. Another way of passing to the semiclassical description
follows from the form of the Hamiltonian (\ref{H}) where the
electron degrees of freedom are exactly eliminated at the cost of
the strongly nonlinear coupling of the phonon degrees of freedom.
The semiclassical decoupling  then means the decoupling of the
vibrons 1 and 2. The classical equations of motions in this variant
are implied by the following classical Hamiltonian as a function of
the coordinates $\gamma_i$ and respective momenta $\pi_i$:

\begin{equation}
H(\gamma_1,\pi_1,\gamma_2,\pi_2)=\frac{1}{2} \left[ \pi_1^2+\pi_2^2+\gamma_1^2+\gamma_2^2 \right]
+ \alpha \gamma_1 \mp \beta\gamma_2 \exp\left(-2(\gamma_1^2+\pi_1^2)\right)
\label{H:clas}
\end{equation}

The trivial linear stability analysis shows that the stationary
point $(\gamma_1, \gamma_2, \pi_1=\pi_2=0)$ is unstable unless
$\alpha \gg \beta$, $\alpha \ll \beta$ or unless the parameters
($\alpha,\beta$) are located in the small domain near the point
$\alpha=\beta=0$. The crucial characteristics of interest is however
the chaoticity of the phase space, in particular, its fraction $\mu$
occupied by the chaotic trajectories. The formulae  which relate
this quantity to the parameters of the statistics of quantum levels
for both short- and long-range statistics are widely known
\cite{Berry:1984,Seligman:1985a,Seligman:1985b}. We investigated the chaoticity of
the trajectories implied by the Hamiltonian (\ref{H:clas}) and
performed rough estimations of the 'chaoticity parameter' $\mu$ for
different energies. We studied the sensitivity of the classical
equations of motion to the initial conditions checking whether two
initially neighboring trajectories diverge in the course of time
(that is studying the Lyapunov index
\cite{Seligman:1985a,Seligman:1985b,Benettin}). The fraction of the
chaotic trajectories at given energy $E$  was then estimated by
taking at random the initial points respecting $H(\gamma_1, \pi_1,
\gamma_2,\pi_2)=E$. The detailed quantitative presentation of this
work is to be given elsewhere; for the sake of the present contribution
we just mention the results briefly. The classical phase space in
the energy range corresponding to the energies of quantum levels
used (that is up to $E\sim 30$) appears to be pronouncedly chaotic.
The fraction of chaotic levels increases rapidly with the increase
of energy and reaches its maximum $\mu\sim 0.90-0.99$ for the
energies typically $E\sim 10$ which roughly corresponds to the lower
energy bin of figure \ref{fig:4} (levels 100-600). With further increasing
energy the value of $\mu$ then slowly decreases. Its estimations for $E$ between
20-30 range from $0.3$ to $0.75$ strongly depending on the parameters
$\alpha$ and $\beta$. This latter energy interval approximately
corresponds to the higher energy bin of figure \ref{fig:4}b. We note once again
the more pronounced universality of the quantum characteristics at
the upper bin (figure \ref{fig:4}b) which appears to correspond to the moderate
values of the degree of the classical chaoticity. The quantitative
conclusions for this item is however a challenge for a separate
study.

In any case, the above results confirm the nonuniversal degree of the
classical nonintegrability demonstrated by the dependence of the
chaoticity parameter $\mu$ on the model parameters $\alpha $ and
$\beta$ and on the location of the energy segment in the spectra.
Moreover, the linear dependence of the spectral rigidities in
certain parts of the spectra implies a strong suppression of the
level repulsions because of the presence of regions with various
degree of chaoticity and related chaos-assisted tunneling between
the regions of small degree of chaoticity.
 Analogous nonuniversal behaviour of the level
compressibilities presented in figure \ref{fig:4} indicates a cumulation of most
of their values in the chaotic region $(0, 0.033)$ close to the
semi-Poisson values $0.5/15$ depending on the spectral segment where
the chaoticity parameter is correspondingly reduced.

\section{Fractal dimensions and multifractality of the JT-excited wavefunctions }

An independent quantitative analysis of the properties of the
E$\otimes (b_1+b_2)$ excited spectra can be obtained by exploring
the scaling of the inverse participation ratio (\ref{D}) and
(\ref{IPR}) in the spectral representation of Fock states and
following calculation of related fractal dimensions. In figure 6, we
give the samples of the scaling of $\log I_{2,L}(n)$ (from equation
(\ref{IPR})) for several successive states as functions of the box
size $l$ ($L\sim 1/l^2$) comprising $l^2$ Fock states. The linear
slope in the log-log coordinates stretches up to the box sizes
8$\times$8 which indicates the existence of a well determined
quantity $D_2$ for each level, although the generalized fractal
dimensions show slight level-to-level fluctuations.

\begin{figure}[h]
\includegraphics[width=0.7\hsize]{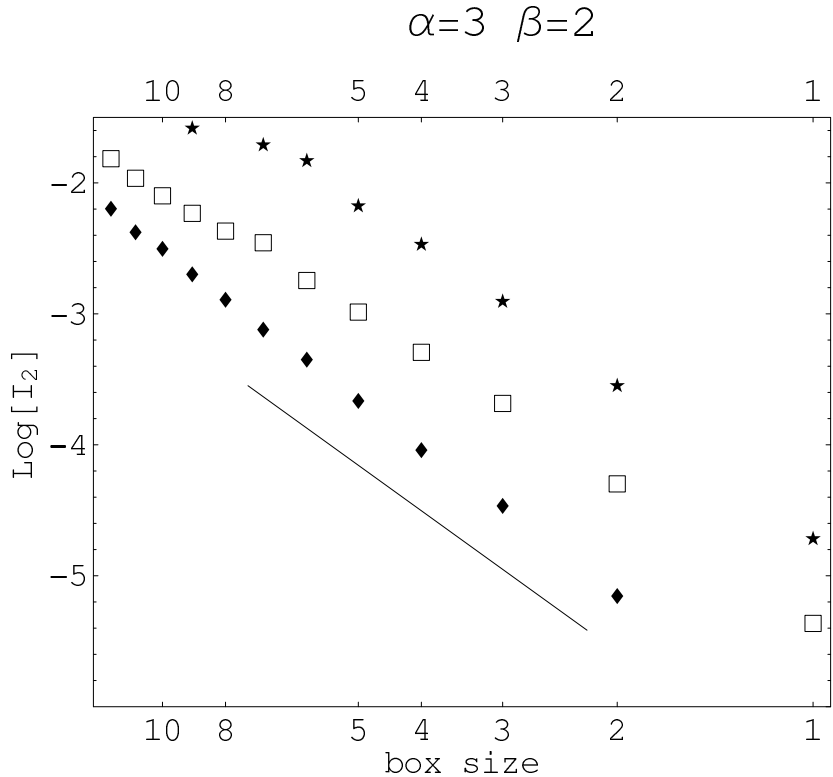}
\caption{Examples of scaling of the inverse
participation ratio $I_{2 l}(n)$ as a function of the box size $l$ in
the box-counting algorithm as the illustration of the reliability of
the scaling (\ref{IPR}) for our model. Different shapes of symbols
correspond to the values for different levels $n$: diamonds for
$n=153$, stars for $n=157$, squares for $n=161$. The domain of linear
slopes (proportional to the fractal dimensions $D_2$) extends up to
the box sizes  $\sim 8\times 8$ (at larger box sizes the linearity
is violated due to the size effects). Slight level-to-level
fluctuations of the slopes are also seen. Similar scaling can be
demonstrated for $q=3,4$ as well.}
\label{fig:6}
\end{figure}

\begin{figure}[h]
\includegraphics[width=0.7\hsize]
{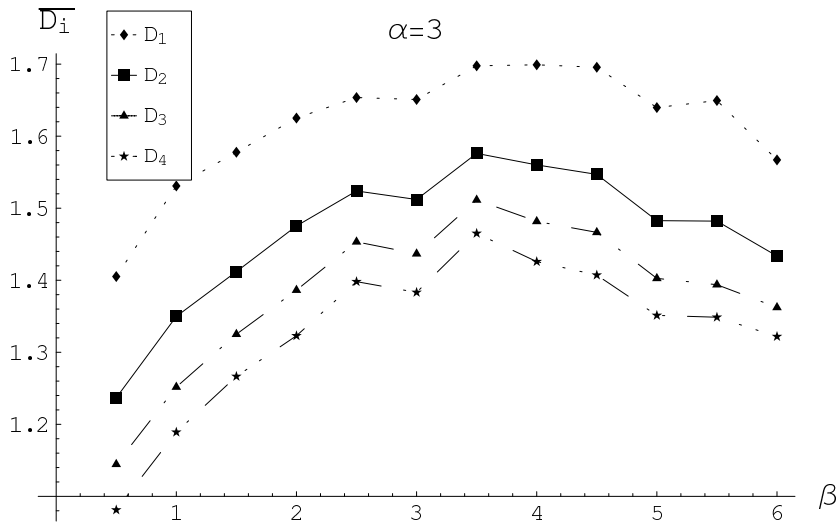} \caption{Fractal dimensions $\bar{D_q}$ averaged
over $n$ for different pairs of $\alpha=3, \beta$.  A weak
 multifractality is evident when comparing $\bar{D_q}$ for different $q$.}
\label{fig:7}
\end{figure}

\begin{figure}[ht]
\includegraphics[width=0.8\hsize]{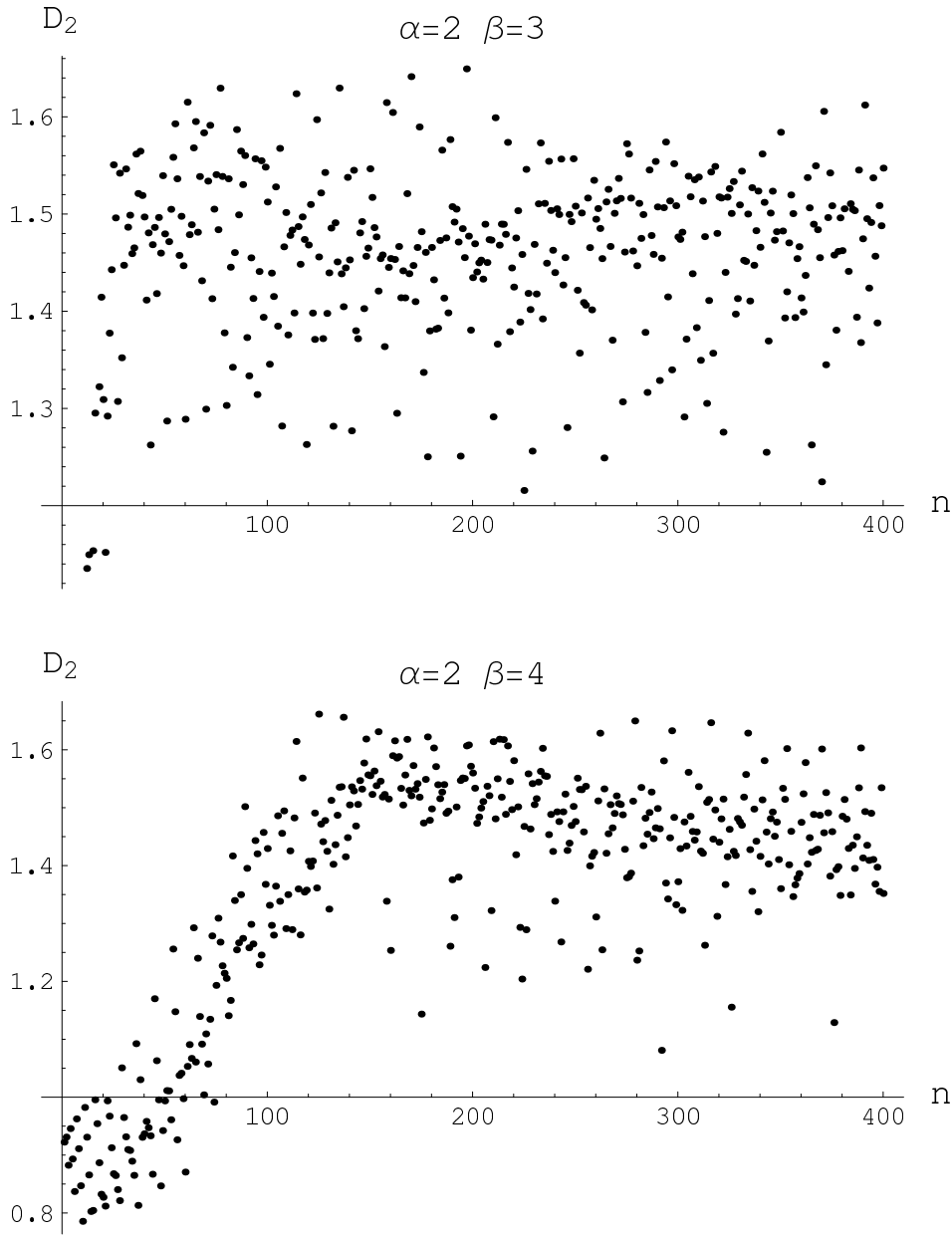}
\caption{Fluctuations of fractal dimensions $D_2$ over levels $n$
for different pairs $\alpha, \beta$.}
\label{fig:8}
\end{figure}

In figure \ref{fig:7}, the averaged fractal dimensions $\bar{D_q}$, $q=1,2,3,4$,
in the spectral representation are displayed for a set of
interaction strengths.
 Small variations of  $\bar{D_q}$ ($\bar{D_i} < \bar{D_j}$ at $i>j$)
testify a weak multifractality of the respective wavefunctions.
The fractal dimensions $D_q$ show strong fluctuations with changing
$n$ (figure \ref{fig:8} so that one has to speak rather about their statistical
distributions \cite{Parshin:1999}. It is seen that this distribution
may have a pronounced drift --  there occurs a crossover between
dimensions $d=1$ and $d=2$ at low $n$. It becomes more homogeneous
when $n$ increases. In view of the interpretation in terms of the 1D
'pseudolattice' in $j$-space with natural size of the order $\sim
\sqrt{n}$ this can be understood as a tendency towards a universal
distribution in the limit of a very large lattice size. Such a
limiting behaviour of statistical characteristics of the
distribution of fractal dimensions for large lattice sizes of the
Anderson model at M-I transition was conjectured by Parshin et al
\cite{Parshin:1999}.

\begin{figure}[hb]
\includegraphics[width=0.75\hsize]
{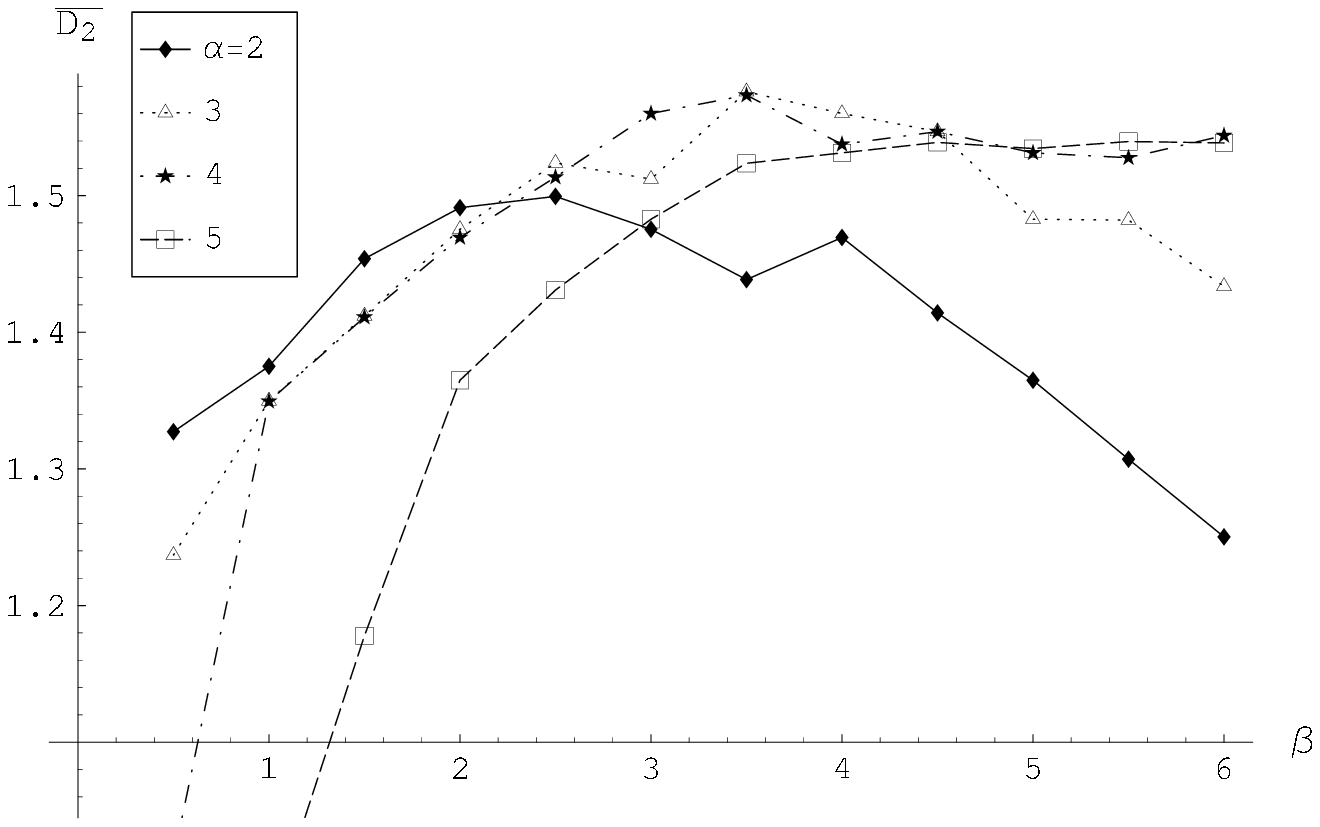} \caption{Fractal dimensions $\bar{D_2}$ averaged
over $n$ in the range 200--400 for different pairs $\alpha, \beta$.
In the 'most chaotic' domain the values of $\bar{D_2}$ are
distributed in a narrow region $\sim 1.5\pm 0.1$.}
\label{fig:9}
\end{figure}

 Despite the level-to-level fluctuations of the fractal dimensions
(figure \ref{fig:8}) their values averaged over level numbers appear to exhibit
marginal universality in a similar sense as the universality marked
in figure \ref{fig:4}. The averaged fractal dimensions $\bar{D_2}$ in figure \ref{fig:9}
show a markable tendency to values $1.5 \pm 0.1$
 apart from the case $\alpha \gg \beta$ or $\alpha \ll \beta$.

\section{Conclusion remarks}

The statistical properties of the investigated JT model are closely
related to its spatial symmetry given by the interaction strengths
$\alpha, \beta $. Our numerical analysis of the spectral rigidity,
its slopes and fractal dimensions bring the evidence for their
apparent nonuniversality and spectral inhomogeneity. Namely, our
results allow us to conclude that the nonuniversality is related  by
the dimensional crossover between $d=1$ and $d=2$ when changing the
parameters $\alpha, \beta$. A support for this suggestion is the
behaviour of the fractal dimension $\bar{D_2}$ (figure \ref{fig:9}) which
approach  very close to the  value $1.5$ at the points $\alpha=\beta
$. In the neighbourhood of this point the nonuniversality of
$\bar{D_2}$ is the most moderate; the deviations from $1.5$  extend
within the interval $\pm 0.1$.

We have shown that although the long-range correlation measure
$\bar{\Delta}_3(\bar n)$ and the partial compressibilities (averaged
over partial segments of the spectra) are nonuniversal and highly
inhomogeneous, the slopes  {\it averaged over the whole relevant
spectra} approach marginally (quantitatively close) to the universal
value characteristic for the semi-Poisson distribution, $0.5/15\sim
0.033$. This behaviour is analogous to the well defined marginal
behaviour of the average fractal dimensions $\bar D_2$ of the
respective wavefunctions mentioned above. To our knowledge, till now
the theory justifying the existence of the limiting 'universal'
value related to the said distribution is lacking. The only relation
$\chi=(d-D_2)/2d$ derived by Kravtsov and Lerner
\cite{Kravtsov:1995} was the first bridge between the fractal
dimension $D_2$ and the long-range level statistics. However, it is
proven to be valid only for small values of compressibility and does
not fit to our calculations.

We have also revealed the corresponding nonuniversal and
inhomogeneous behaviour of the chaoticity parameter - the fraction
of the chaotic phase space of the classical trajectories. The
detailed analysis of this parameter in the whole phase space would
be very useful for further considerations of phenomena especially
related to the mixing \cite{Ketzmerick:1996} of the phase space
between the regions of different degrees of chaoticity. The
presence of the mixing phenomena is supported by the multifractal
behaviour of the wavefunctions. Because of the nonexistence of the
extended states in our model we can conclude on the concept of the
chaos-assisted tunnelling between the domains with reduced 
stochasticity through those with high degree of stochasticity.

Besides the Anderson model close
 to the metal-insulator transition \cite{Shklovskii:93} another models sharing
 the universal semi-Poisson statistics are known - in particular,
 the Bogomolny \cite{Bogomolny:1999} plasma model with screened
Coulomb interactions  and the 'daisy models' \cite{Hernandez:1999}.

\ack We thank Dr. P. Marko\v s for useful discussions.
  The support of the project by the Grant Agency of the Czech Republic
 No. 202/06/0396 is greatly acknowledged. Partial support is acknowledged also
 for the project No. 2/6073/26 by the Grant Agency VEGA, Bratislava.

\end{document}